\documentclass[prl,twocolumn,superscriptaddress]{revtex4-1}
\usepackage{graphicx}
\usepackage{gensymb}
\usepackage{amsmath}
\usepackage{amssymb}
\usepackage{mathrsfs}
\usepackage{xcolor}
\usepackage{multirow}
\usepackage[normalem]{ulem}
\usepackage{color}
\usepackage{cancel}
\usepackage{graphicx}
\usepackage{dcolumn}
\usepackage{bm}
\begin{document}


\title{Swirling and snaking, 3D oscillatory bifurcations of vesicle dynamics in microcirculation}


\author{Jinming Lyu}
\affiliation{Univ. Grenoble Alpes, CNRS, Grenoble INP, LRP, Grenoble, France }

\author{Paul G. Chen}
\affiliation{Aix Marseille Univ, CNRS, Centrale Marseille, M2P2, Marseille, France}

\author{Alexander Farutin}
\affiliation{Univ. Grenoble Alpes, CNRS, LIPhy, Grenoble, France }

\author{Marc Jaeger}
\affiliation{Aix Marseille Univ, CNRS, Centrale Marseille, M2P2, Marseille, France}

\author{Chaouqi Misbah}
\affiliation{Univ. Grenoble Alpes, CNRS, LIPhy, Grenoble, France }

\author{Marc Leonetti}
\email[]{marc.leonetti@univ-grenoble-alpes.fr}
\affiliation{Univ. Grenoble Alpes, CNRS, Grenoble INP, LRP, Grenoble, France }


\date{\today}

\begin{abstract}
Vesicles are soft elastic bodies with distinctive mechanical properties such as bending resistance, membrane fluidity, and their strong ability to deform, mimicking some properties of biological cells. While previous three-dimensional (3D) studies have identified stationary  shapes such as slipper and axisymmetric ones, we report a complete phase diagram of 3D vesicle dynamics in a bounded Poiseuille flow with two more oscillatory dynamics, 3D snaking and swirling. 3D snaking is characterized by planar oscillatory motion of the mass center and shape deformations, which is unstable and leads to swirling or slipper. Swirling emerges from supercritical pitchfork bifurcation. The mass center moves along a helix, the preserved shape rolls on itself and spins around the flow direction. Swirling can coexist with slipper. 
\end{abstract}

\pacs{}
\maketitle

Soft particles such as capsules, vesicles, red blood cells, compound droplets, and elastic fibers display rich dynamic behaviors in linear and quadratic flows \cite{Vlahovska2009CRAS,Roure2019ARFM,ZhuGallaire2017PRL,Slowicka2013EPJE}. If the zoology of dynamics shares some common characteristics, the so-called tank-treading motion of the interface, for example, can also differ by bifurcation dynamics of each soft particle. Indeed, their shapes and their dynamics depend on the nonlinear coupling between hydrodynamic stresses and interfacial mechanical response. While elastic resistance to shear and stretching governs the response of the solid membrane of capsules, bending resistance dictates one of the fluid membranes of vesicles.

If deciphering the complex dynamics of such deformable particles is still an open issue, the challenge is also to measure some physical/mechanical membrane properties by an inverse method comparing experimental shapes to numerical ones \cite{LinSui2021PoF,Lefebvre2008PoF,Tregouet2018PRF,Loubens2015JFM,Salipante2021SM,Lin2021SM}. In the high-throughput shape recognition of red blood cells moving in a microcapillary, it is essential to select relevant physical parameters (e.g., flow rate and confinement) to avoid any oscillatory or transient dynamics for efficient shape recognition \cite{Kihm2018Plos,Guckenberger2018SM}. 

The motion of active particles is also a guide to infer new dynamics. Indeed, the bacterial pathogen listeria propels itself along helical trajectories, for example \cite{Zeile2005,Shenoy2007}. Is it possible with a passive soft particle ? 

In this Letter, we focus on the stability of a confined three-dimensional vesicle (position of the mass center (CM) and shape) in a \textit{bounded} Poiseuille flow, a different configuration from the \textit{unbounded} one where the vesicle is free to move away from the centerline. By considering several kinds of perturbations of a vesicle's position, we formulate the following questions: Does the snaking motion observed in two dimensions exist in three dimensions despite the symmetry of round microcapillaries? Can we expect novel three-dimensional oscillatory dynamics ? If so, how to characterize the transitions (limit cycles?) and their potential coexistence with stationary shapes (fixed points) such as axisymmetrical ones (bullets and parachutes) and slippers observed in bounded and unbounded Poiseuille flows ?

\begin{figure}[h!]
\begin{center}
\includegraphics[trim=16 2 10 20, clip,width=8cm]{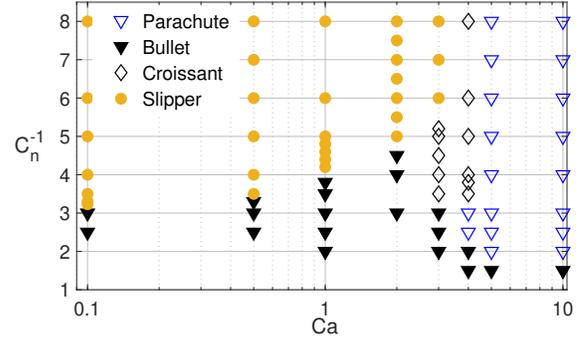}
\end{center}
\caption{Phase diagram of moderately deflated vesicle ($v$$=$$0.95$) moving in a microcapillary. All the shapes are stationary. The diagram is similar for $v$$=$$0.9$.}
\label{fig:fig1}
\end{figure}

\begin{figure*}
\includegraphics[trim=0 0 0 0, clip,width=12.5cm]{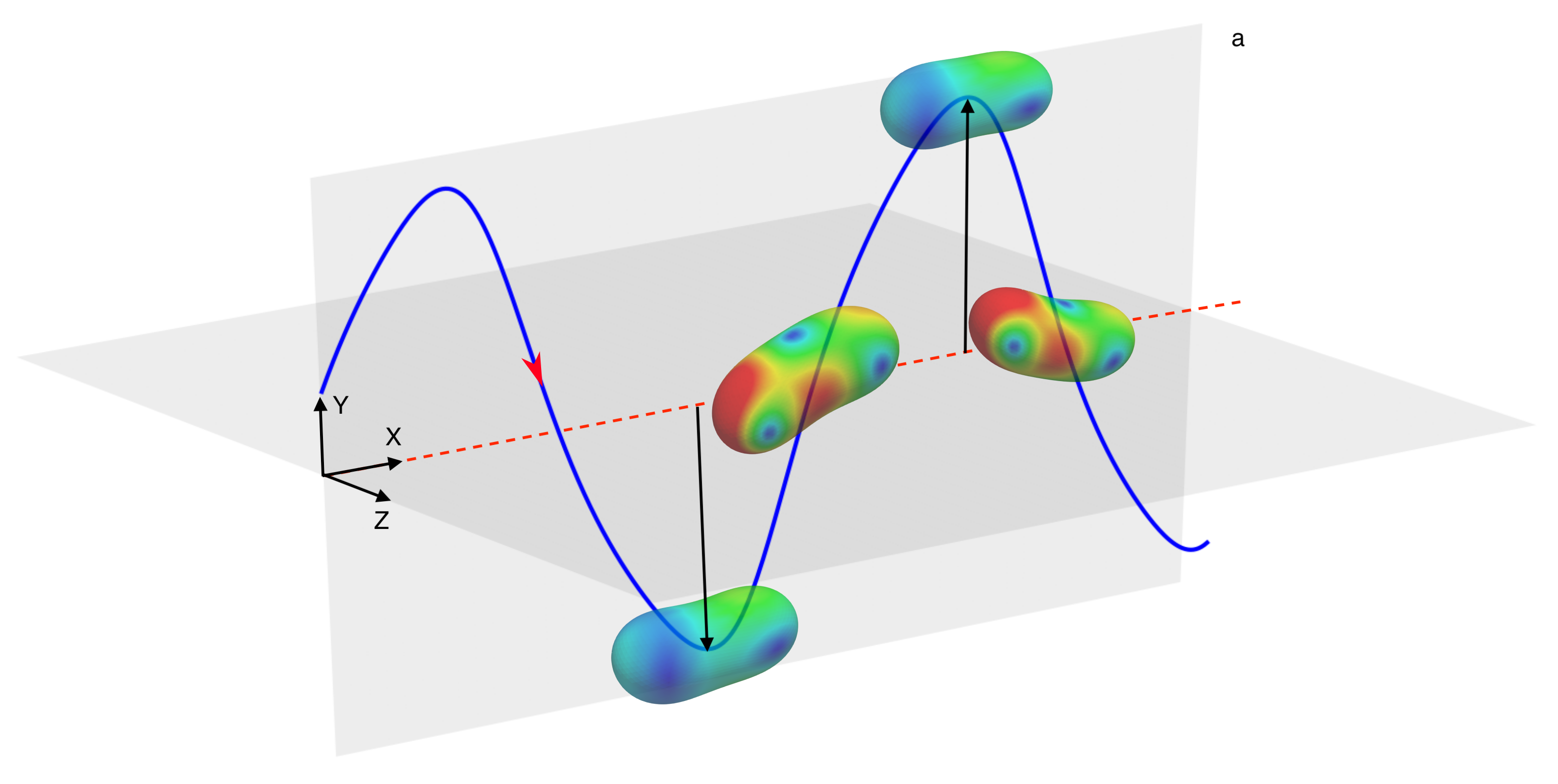}\\\includegraphics[trim=0 0 0 0, clip,width=12cm]{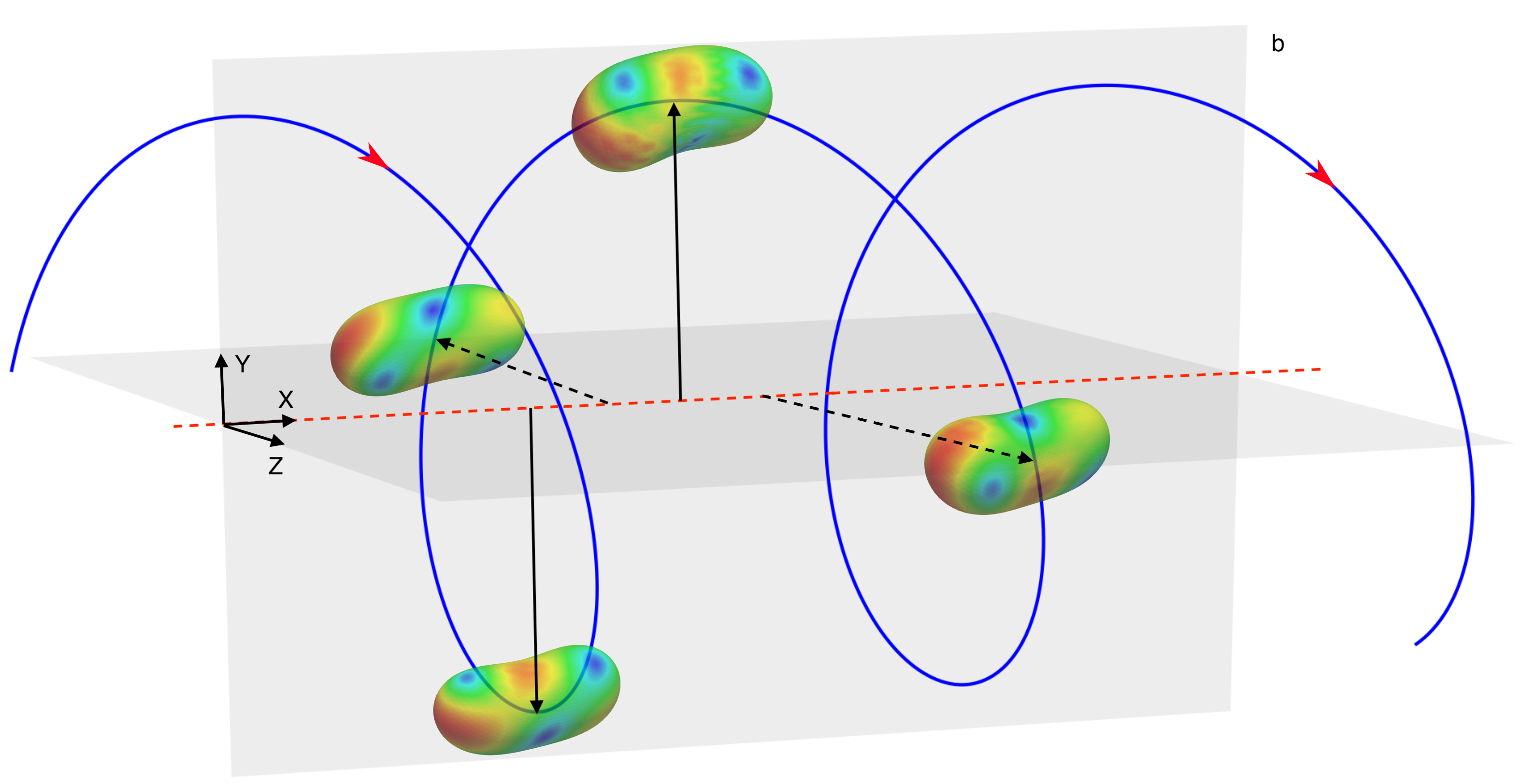}
\caption{3D snaking and swirling: $v$$=$$0.85$, $Ca = 1$ and $C_n = 0.364$. a - 3D Snaking: $\theta_y(0) = 0$. The center of mass oscillates in the plane (x,y): blue curve. The shape deforms. b - Swirling: $\theta_y(0) \neq 0$. The center of mass moves along a helix (blue curve) centered on the microcapillary's axis, the $x$-axis. The shape does not deform and turns around the helix. The black arrows point out to the center of mass with solid lines in the plane (x,y) and dashed lines in the plane (x,z). In a and b, the radial positions of the center of mass are overestimated compared to the size of vesicles for a better visualisation. Color code corresponds to interfacial velocity: see supplementary information for videos.}
\label{fig:fig2}
\end{figure*}

We consider a vesicle of volume $V=\frac{4}{3}\pi R_0^3$ and area $A$ in a \textit{bounded} Poiseuille flow, i.e., in a round microcapillary of radius $R_c$,
\begin{equation}
\textbf{V}_\infty(\textbf{x})\,=\,V_m\big(1-r^2\,/\,R_c^2)\,\textbf{e}_x,\quad r \leq R_c
\end{equation}
where $r^2=y^2+z^2$, $\textbf{x}=(x,y,z)$ and $V_m$ the velocity at the centerline. 
The vesicle's dynamics and shape in microcirculation are determined by three dimensionless parameters: the reduced volume $v$ (a measure of the vesicle's deflation), the capillary number $Ca$ (the ratio of viscous to bending forces), and the confinement $C_n$ (the ratio of the vesicle effective radius to the capillary radius):
\begin{equation}
v\,=\,\frac{V}{\frac{4\pi}{3}(\frac{A}{4\pi})^{3/2}}\,\,;\,\,C_n=\frac{R_0}{R_c}\,\,;\,\,Ca\,=\,\frac{\eta V_m R_0^3}{\kappa R_c}
\end{equation}
where $\kappa$ is the membrane bending modulus and $\eta$ the fluid viscosity.
The system is investigated by numerical simulations using a coupled isogeometric finite-element method with boundary-element method reported in \cite{Lyu2021}, which is a direct extension of the previous work \cite{Boedec2017JCP} on soft particles in unbounded Stokes flow. The code has been validated by comparison with \cite{Boedec2011JCP} in free space and with \cite{Trozzo2015JCP} in confined configuration. The simulation results in the limit of strong confinement are also in quantitative agreement with \cite{Barakat2018JFMa,Barakat2018JFMb,Chen2020PRF}. As underlined in \cite{Barakat2018JFMb,Agarwal2020PRF}, 3D computations of vesicles in \textit{bounded} configuration is still a challenge contrary to capsules. To limit the parameter space, we fix the viscosity contrast of unity, and computations are performed by decreasing the reduced volume by step $\Delta v=0.05$. When an oscillatory bifurcation is observed, the reduced volume is no longer decreased, and the bifurcation is analyzed. In what follows (text and figures), length, time, and pressure are made dimensionless by $R_0$, $\eta R_0^3/\kappa$, and $\kappa/R_0^3$, respectively. The initial shape is the solution at the thermodynamic equilibrium. To investigate the stability of the known axisymmetric solution and identify the branches of solutions, the mass center is first moved along the y-axis: $Y_{CM}$$(t$$=$$0)$$=$$H\geq0.0005)$. Then, the longest axis is also turned around the y-axis of an angle $\theta_y(0)$.

\begin{figure*}
\includegraphics[trim=10 0 10 0, clip,width=18cm]{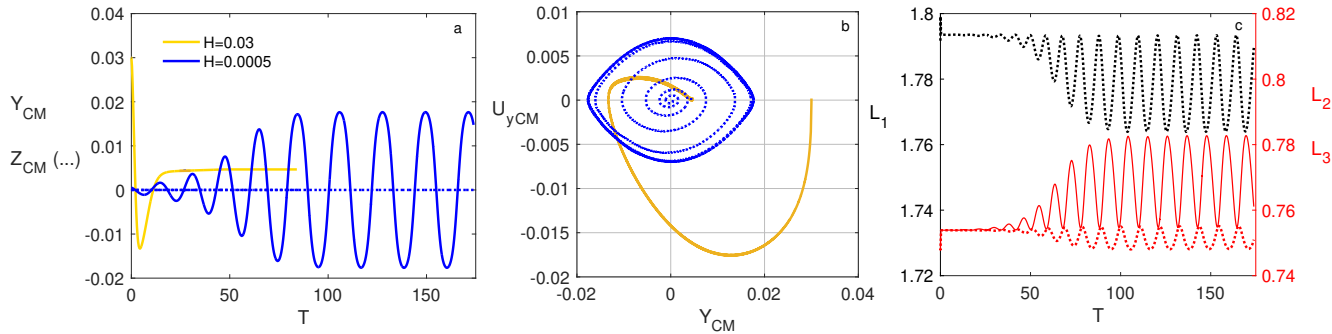}
\caption{3D Snaking: $v$$=$$0.85$, $Ca$$=$$2.3$ and $C_n$$=$$0.4$. (a) The same initial shape evolves to slipper (orange, $H$$=$$0.03$) or snaking (blue, $H$$=$$0.0005$) following the initial position $H$. Snaking is an oscillation of the mass center position in a plane, here ($x,y$). Slipper and snaking can coexist. (b) Phase portrait of (a) showing fixed point (slipper) and limit cycle (snaking). (c) The 3D shape oscillates during snaking: temporal variations of the lengths $L_j$ of the three semi-axis of the equivalent ellipsoid.}
\label{fig:fig3}
\end{figure*}

\begin{figure}
\includegraphics[trim=25 0 0 0, clip,width=7cm]{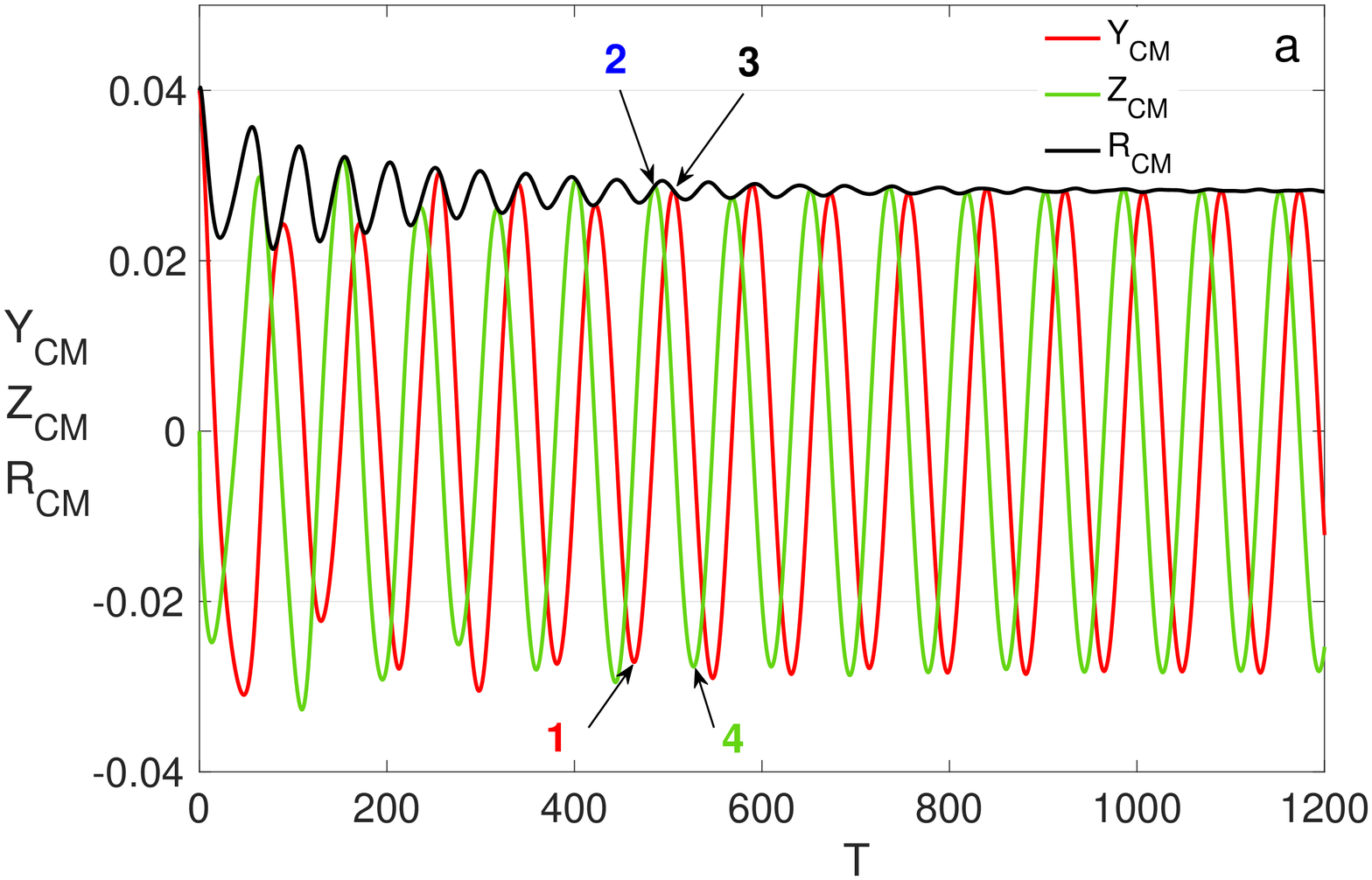}\\\includegraphics[trim=20 0 0 0, clip,width=6.9cm]{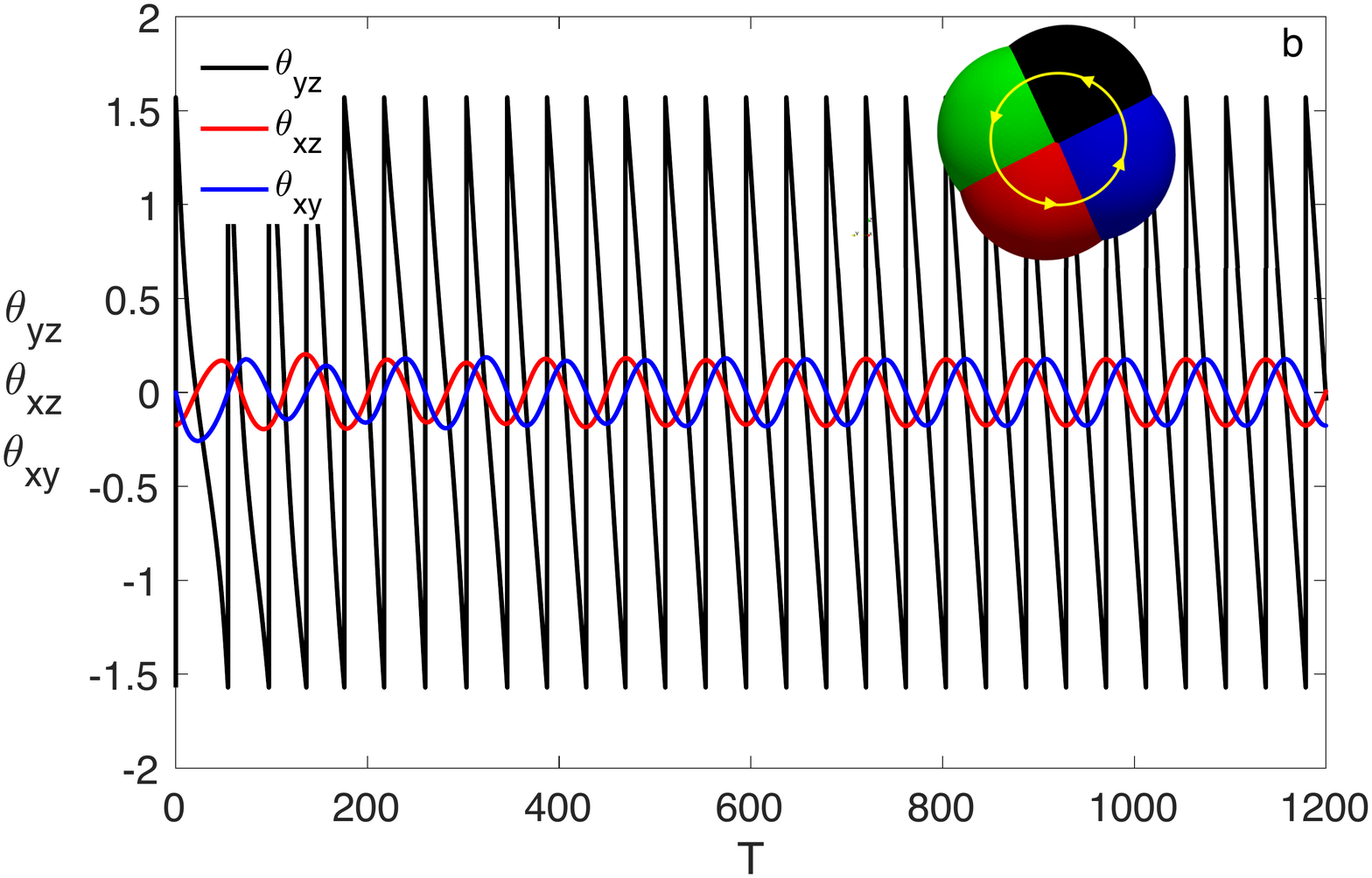}\\\includegraphics[trim=0 10 0 10, clip,width=7cm]{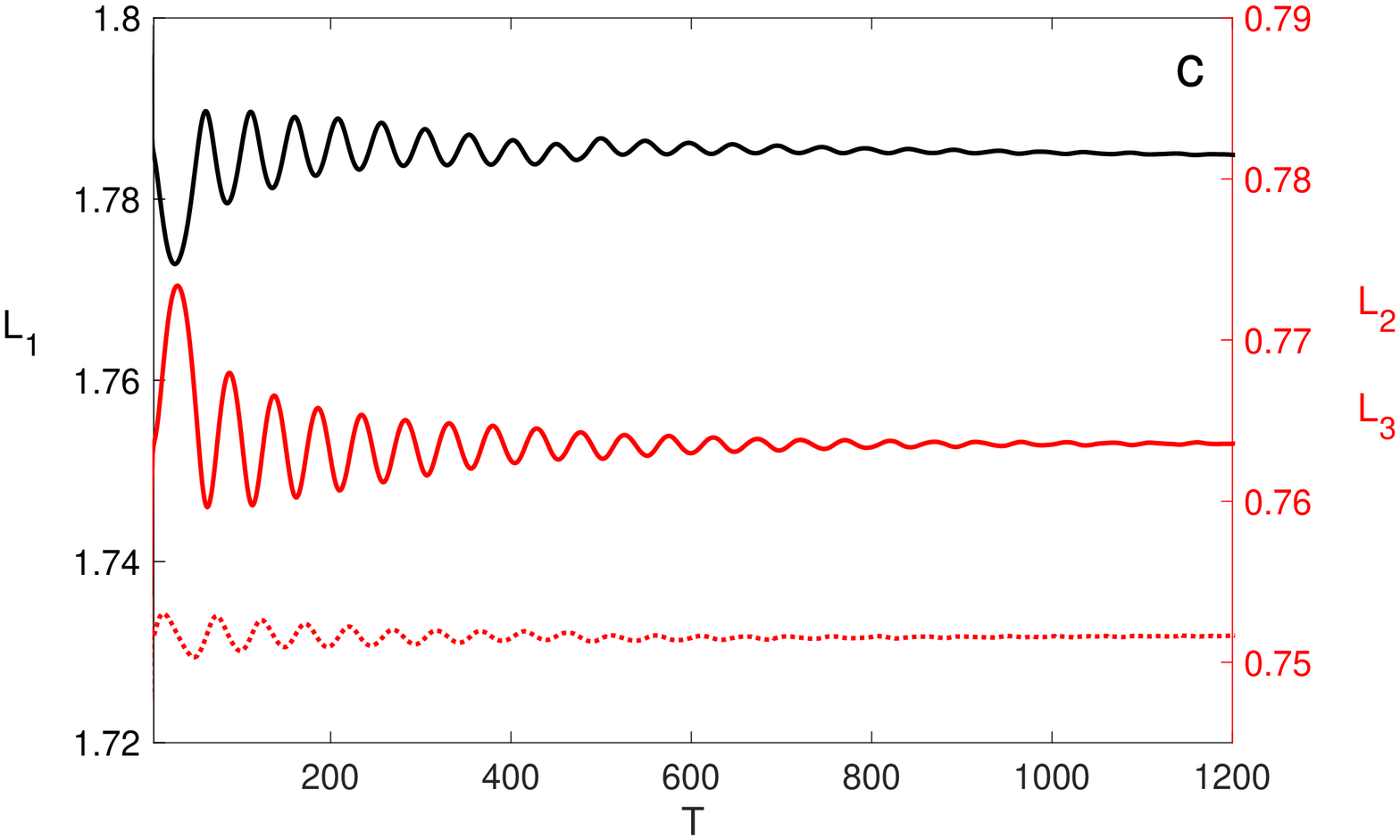}
\caption{Swirling: $v$$=$$0.85$, $Ca$$=$$1$ and $C_n$$=$$0.364$. Here, the initial state is the snaking shape ($Y_{CM}\approx 0.04$, the maximum) which is turned of $\theta_y$$=$$10^{\circ}$ around the $y$-axis. a - Temporal variations of the cartesian and radial position of the mass center $(Y_{CM},Z_{CM},R_{CM})$ where $R_{CM}$$=$$\sqrt{Y_{CM}^2+Z_{CM}^2}$. b - The angles $\theta_{xy}$ and $\theta_{xz}$ of the longest semi-axis $L_1$ oscillate in quadrature in the planes (x,y) and (x,z) respectively. The angle $\theta_{yz}$ varies linearly with time. The insert refers to the shape and its rotation (yellow arrow) during one period seen from the rear at four times 1-4 defined in a: 1-red, 2-blue, 3-black, and 4-green. See movie 2 in supplemental materials. c - The lengths $L_j$ of three semi-axis of the equivalent ellipsoid tend to a constant contrary to snaking (Fig.\ref{fig:fig3}-c). }
\label{fig:fig4}
\end{figure}

\begin{figure}
\includegraphics[trim=0 140 0 140, clip,width=4cm]{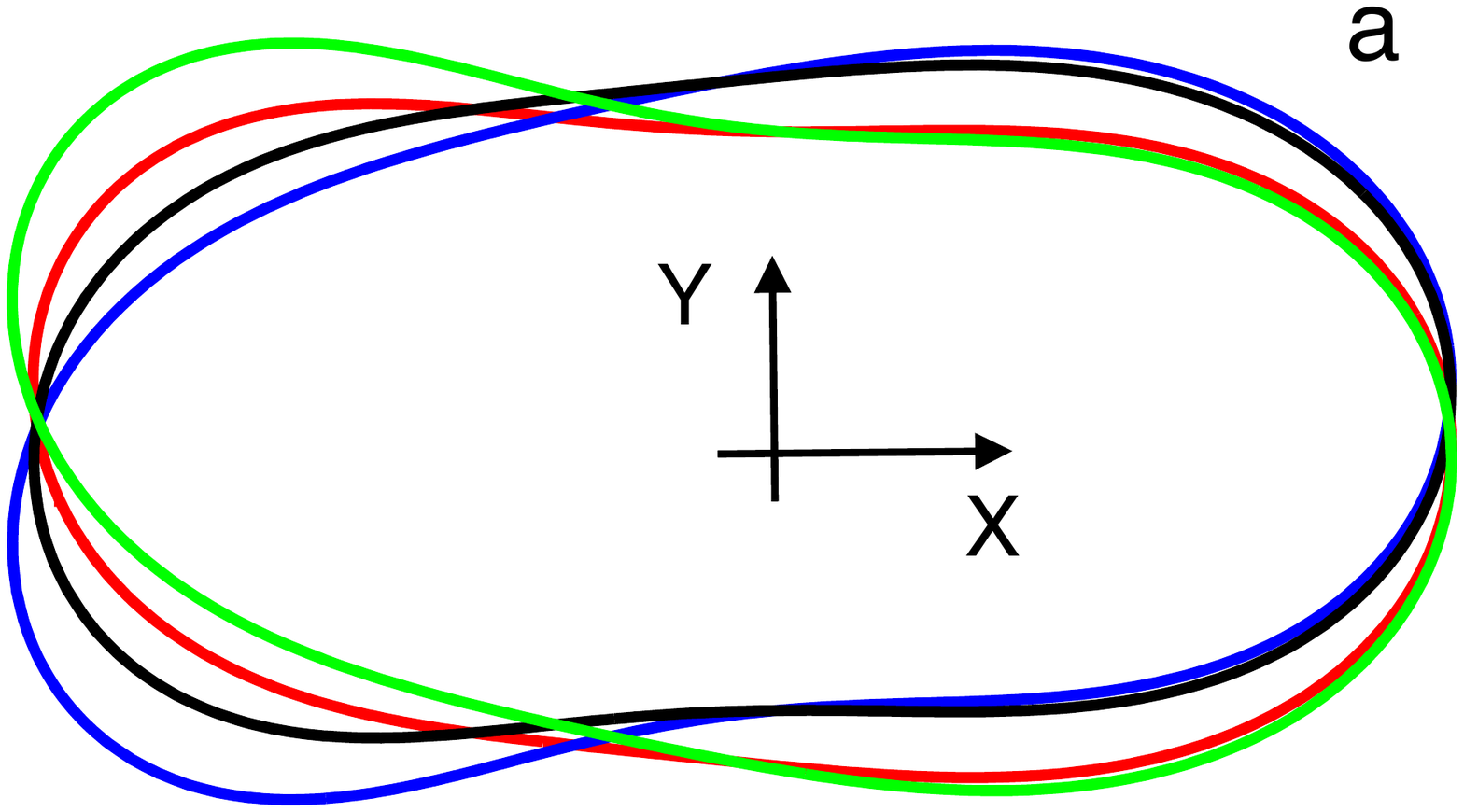}\includegraphics[trim=0 140 0 140, clip,width=4cm]{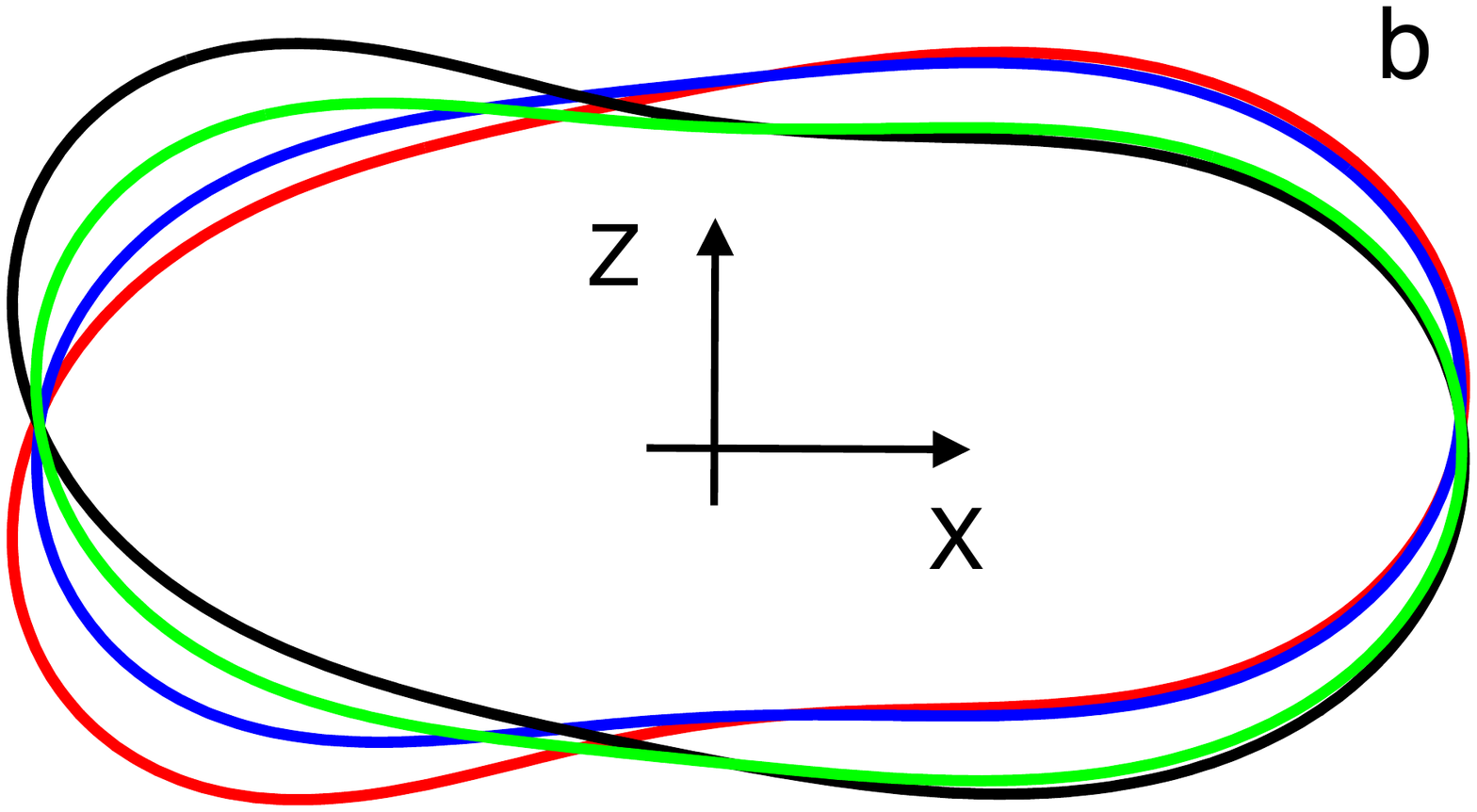}
\caption{Swirling: $v$$=$$0.85$, $Ca$$=$$1$ and  $C_n$$=$$0.364$. a - Cross-sections of the shapes in the plane $(x,y)$ at the times 1-4 defined in Fig.\ref{fig:fig4}-a. b - Cross-sections of the shapes in the plane $(x,z)$. See text for details on the symmetries of shapes.}
\label{fig:fig5}
\end{figure}

\textit{Vesicles with moderate reduced volume: $v \geq 0.9$}. Two values were studied, namely  $v=0.9$ and $v=0.95$. Whatever the initial perturbation [$H$ and $\theta_y(0)$], the capillary number $Ca$ and the confinement $C_n$, all the final states are stationary, as shown in Fig.\ref{fig:fig1}. Bullet and parachute shapes are axisymmetric with a positive and negative curvature at the rear respectively \cite{Coupier2012PRL}. Croissant-like shapes are symmetric to $(y,z)$ and $(x,z)$ planes or any other couple obtained by rotation of these planes around the $x$-axis. This mode is out of the scope of this study and won't be evoked further.  The mass center of slipper shapes is out-of-axis with one plane of symmetry, $(x,y)$ for example. The phase diagram is qualitatively similar for $v = 0.9$ with a shift of slipper-croissant-parachute transitions to larger capillary numbers. Slippers were first observed in 2D \cite{Kaoui2009PRL}, then in the 3D unbounded case \cite{Farutin2014PRE1} and finally very recently in the 3D bounded case \cite{Agarwal2020PRF,Barakat2018JFMb}. Our results are in qualitative agreement with those of \cite{Agarwal2020PRF} and quantitative agreement with \cite{Farutin2014PRE1} in the \textit{unbounded} limit. The axisymmetric/slipper transition is a supercritical pitchfork bifurcation with the capillary number and the confinement in the limit of the accuracy of our simulations (data not shown).

\textit{Vesicles with lower reduced volume: $\theta_y(0)=0$}. We consider $v=0.85$. In strong confinement $C_n \geq 0.8$, whatever the initial position $H$ and the capillary number, the final state is stationary and axisymmetric as observed in experiments \cite{Vitkova2004EPL,Coupier2012PRL}. As for the droplets \cite{Bretherton1961}, the pressure in the thin film between membrane and wall imposes the axisymmetry by relaxing any shape deformation \cite{Trozzo2015JCP,Barakat2018JFMa,Barakat2018JFMb,Chen2020PRF}. But, contrary to a droplet, there is no inner recirculation. When confinement becomes smaller (larger channel's radius), any deviation from the centerline leads to an additional dissipation due to membrane tank-treading and so inner recirculation. At the same time, the shape can deform where the shear is highest leading to a slower velocity along the microcapillary and less dissipation \cite{Kaoui2009PRL}.  This competition exists in unbounded and bounded flows showing that the role of the wall takes part further. The higher the mass center to the centerline, the stronger the repulsive hydrodynamic contribution of the wall, also called viscous lift force \cite{Olla1997,Sukumaran2001PRE,Abkarian2002PRL,Callens2008EPL,Zhao2011PoF}. Thus, when the vesicle leaves too much the centerline, the deformation is damped, the vesicle pulls back. As the shape and the microcapillary are symmetric to the plane $(x,y)$, the repulsive wall's force is also radial leading to a 2D oscillation of the mass center in $(x,y)$ as shown in Fig.\ref{fig:fig2}-a and Fig.\ref{fig:fig3}-a. Snaking is characterized by a limit cycle, slipper by a fixed point as shown in Fig.\ref{fig:fig3}-b. This 3D oscillatory dynamics where shape's deformations are involved (Fig.\ref{fig:fig3}-c) is called snaking by analogy with the results in two dimensions \cite{Kaoui2011complexity}: see movie 1 in the supplemental material. As we will see further, snaking is a supercritical pitchfork bifurcation characterized by a threshold and a continuous transition. In weaker confinement, the configuration where slippers appear is recovered as in literature. At the transition snaking-slipper, snaking appears if the initial condition $H$ is smaller than a critical value. Otherwise, slipper appears highlighting a domain of coexistence. The final state can be characterized by $Y_{CM}<A_{CM}^{snaking}$ as shown in Fig.\ref{fig:fig3}-a. $A_{CM}^{snaking}$ decreases with the capillary number. To the best of our knowledge, this is the first observation and characterization of snaking in 3D. Its stability will be discussed further.

\textit{Vesicles with lower reduced volume: $\theta_y(0)\neq 0$}. Snaking is obtained when the longest axis is initially in the flow direction ($x$-axis). What happens if the symmetry is broken by rotating the longest axis of an angle $\theta_y(0)\neq 0$ around the $y$-axis? Initially, $H$$=$$Y_{CM}(0)$$\neq$$0$ and $Z_{CM}(0)$$=$$0$. Consider an example ($Ca$$=$$1$, $C_n$$=$$0.364$, and $\theta_y(0)$$=$$10^{\circ}$) where the initial shape is the snaking one when the mass center is maximum: $Y_{CM}\approx0.04$. As shown in Fig.\ref{fig:fig4}-a, after a transient, $Z_{CM}$ and $Y_{CM}$ oscillate in quadrature with the same amplitude. The mean position of the mass center is on the x-axis: $<Y_{CM}(t)>=<Z_{CM}(t)>=0$. Thus, the mass center moves along a helix of microcapillary's axis ($x$-axis) of radius $R_{CM}$$=$$\sqrt{Y_{CM}^2+Z_{CM}^2}$ as shown in Fig.\ref{fig:fig2}-b and Fig. \ref{fig:fig4}-a. We call this new phenomenon swirling. The same dynamics is obtained in the case of a linear perturbation  $H<<1$ and $\theta_y(0)=10^\circ$. However, it is better for accuracy to begin with large initial position $H$ due to unavoidable numerical errors on such a long time of simulations. Close to the centerline, the physical origin is a competition between dissipations induced by on the one hand, membrane tank-treading and on the other hand, shape deformation associated with a lower velocity \cite{Kaoui2009PRL} whatever the orientation. Contrary to snaking, the mirror symmetry is broken, namely the symmetry with respect to the (x,y) plane. Now, the lift force depends on the orientation of the vesicle in the microcapillary and becomes too intricate to decipher quantitatively its role. However, some relevant insights can be gained considering shape and orientation in relation to the invariance along the $x$-axis. Firstly, the three semi-axis lengths are constant during swirling as shown in Fig.\ref{fig:fig4}-c. Note that the temporal variations are strikingly different from snaking (Fig.\ref{fig:fig3}-c). Secondly, the vesicle turns on itself along the helix at a constant angular rate $d\theta_{yz}/dt<0$ in the plane (y,z) as shown in Fig.\ref{fig:fig4}-b and its insert (yellow arrow): see movie 2 in the supplementary information to clearly visualize this self-rotation well identified by the protuberance at the rear.  Thirdly, we check our statements on cross-sections in $(x,y)$ and $(x,z)$ planes along the helix in four points 1-4 defined in Fig.\ref{fig:fig4}-a. Consider times 1 (red) and 2 (blue). The red cross-section in $(x,z)$ ($(x,y)$) becomes the blue one in $(x,y)$ ($(x,z)$ plus the symmetry to the $x$-axis) as shown in Fig.\ref{fig:fig5}. It corresponds to the rotation of $\pi/2$. Consider times 2 and 3 (black). The blue cross-section in $(x,y)$ becomes the black one in $(x,z)$ by the symmetry to the $x$-axis. It corresponds to another rotation of $\pi/2$ and so on. Finally, all the cross-sections can be deduced from both by symmetry with respect to x-axis. Thus, the invariance along the microcapillary imposes swirling dynamics with a characteristic constant shape. Boundary conditions at the membrane enforce membrane recirculation and inner flow, unlike a solid body as shown by the interfacial flow in movie 2. 

\begin{figure}
\includegraphics[trim=10 10 5 20, clip,width=8.8cm]{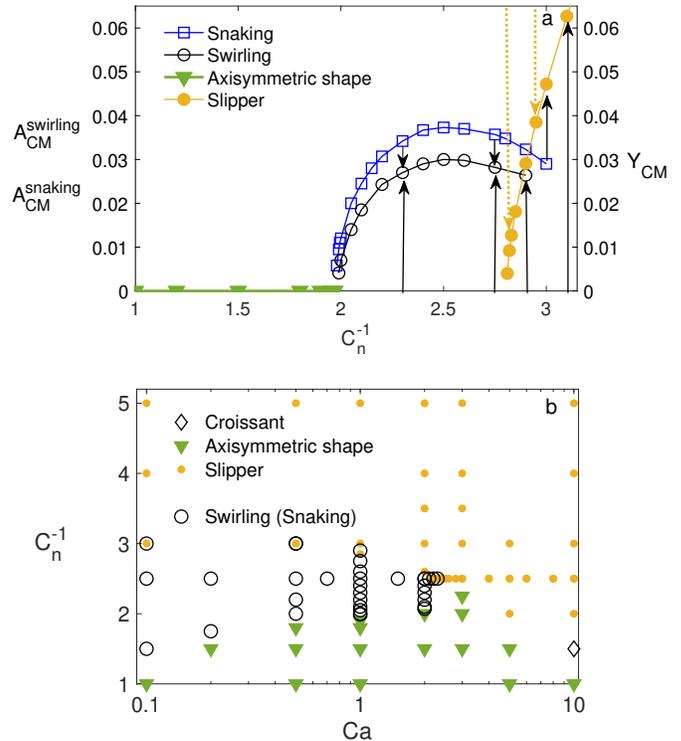}
\caption{Analysis of stability. a - Diagram of bifurcation: $v$$=$$0.85$ and $Ca$$=$$1$. Bifurcations are characterized by the confinement-induced variations of the position of the mass center ($Y_{CM}$) in slipper mode and stable (unstable) stationary axisymmetric shape and the amplitudes ($A_{CM}^{snaking}, A_{CM}^{swirling}$) of the oscillations of the mass center. The arrow at $C_n=0.322$ means that the axisymmetric shape is linearly unstable and evolves to slipper whatever the perturbation. The other black arrows correspond to a rotation of angle $\theta_y=10^\circ$ around the $y$-axis. The cases at $C_n=0.364$ and $C_n=0.435$ show the snaking-swirling transition. The snaking at $C_n=0.333$ is locally stable for a small translation of the shape along $y$-axis but unstable if the shape is turned. There is a domain of coexistence. b - Phase diagram in the plane $(Ca,C_n)$.}
\label{fig:fig6}
\end{figure}

\textit{Analysis of stability: is swirling or snaking the main oscillatory dynamics in Poiseuille flow ?} - The bifurcation diagram has five branches of solutions: stable and unstable axisymmetric shapes, slipper, snaking and swirling as shown in Fig.\ref{fig:fig6}-a. The axisymmetric shape-swirling transition appears as a supercritical pitchfork bifurcation. Indeed, the axisymmetric shape is linearly unstable above a threshold, and the amplitude of oscillations grows continuously. If $\theta_y(0)\neq0$, swirling appears while if $\theta_y(0)=0$, snaking appears. The threshold of snaking and swirling is the same as the accuracy of numerical simulations. This unexpected result can be understood by considering two Hopf bifurcations along the $y$-axis and $z$-axis characterized by two non-linearly coupled amplitude equations. Beyond the mirror symmetry and the one by rotation, the key point is that the physics is the same on both axis allowing the determination of relations between coefficients. It is possible to show that snaking and swirling are two solutions and that snaking is unstable leading to swirling: theoretical details will be published elsewhere. Indeed,  if snaking is numerically stable under small perturbation of the $y$-position of the mass center, it is unstable if a perturbation of orientation ($\approx 10^\circ$ for example) is applied. Due to the high numerical cost of this kind of simulation, only a few perturbations were performed. The smaller the perturbation of orientation ($30$ to $5^\circ$), the longer the snaking-swirling transition time. Thus we cannot exclude that snaking is linearly stable. On the contrary, swirling is stable whatever the perturbation up to the amplitude $A_{CM}^{swirling}$ of swirling oscillations is of the same order of the slipper position $Y_{CM}$. In the small range $0.333 < C_n < 0.344$, the motion of the mass center shows a signature of chaos via a cascade of period doubling during a time longer than that in simulations of Fig.\ref{fig:fig4}. Due to numerical limits, we are unable to claim if this phenomenon is transient or not. Slipper appears also as a supercritical pitchfork bifurcation highlighting a domain of coexistence with snaking and swirling as shown in Fig.\ref{fig:fig6}-a. But, the emerging mode depends on the initial position. Large $H$ (orange arrows) corresponds to slipper, small $H$ to swirling. Finally, the domain of existence of swirling in the parameter space $(Ca,C_n)$ is determined: Fig.\ref{fig:fig6}-b. This is the first observation and nonlinear characterization of swirling. This phenomenon has also been observed in the case $Ca=1$, $C_n=0.4$, $v=0.8$ and $v=0.75$ ensuring its general relevance. 


In summary, the dynamics of vesicles in Stokes flow reveals richer nonlinear dynamics than previously expected. The dynamics of snaking is unstable and leads to slipper or a new oscillatory mode called swirling. One striking difference between snaking and swirling is the role played by the vesicle's deformation; the mass center in swirling moves along a helix without any deformations contrary to snaking.  Thus, helical trajectories are a common dynamics of living systems and biomimetic ones. If the symmetry of a rounded microcapillary imposes swirling, other oscillatory modes might emerge depending on the geometry of channels. In a slit geometry, snaking is expected for example. We hope the present results will aid in shape recognition in microcirculation to prevent biased analysis when comparing experiments and simulations.\\

\begin{acknowledgments}
J. L. wishes to acknowledge the support of LabEx Tec21 (ANR-11-LABX-0030). This work has benefited from financial support from the ANR 2DVISC (ANR-18-CE06-0008-01) and the microgravity research grant of CNES. Centres de Calcul Intensif d'Aix-Marseille and Grenoble are acknowledged for granting access to their high-performance computing resources.
\end{acknowledgments}

\bibliography{article.bib}

\end{document}